\documentclass[reprint,aps,prd,superscriptaddress,showkeys,showpacs]{revtex4-1}
\usepackage{epsfig,amsmath,natbib}

\usepackage{aas_macros}
\usepackage{amssymb}
\usepackage{amsmath}
\usepackage{dsfont}
\usepackage{hyperref}
\usepackage{color}
\usepackage{pbox}
\usepackage{booktabs}
\usepackage[dvipsnames]{xcolor}

\hypersetup{
	colorlinks=false,
	citecolor=green
}

\newcommand{\bb}[1]{\mathbf{#1}}
\newcommand{\bbh}[1]{\mathbf{\hat{#1}}}
\newcommand{\h}[1]{\hat{#1}}
\newcommand\lsim{\mathrel{\rlap{\lower4pt\hbox{\hskip1pt$\sim$}}
        \raise1pt\hbox{$<$}}}
\newcommand\gsim{\mathrel{\rlap{\lower4pt\hbox{\hskip1pt$\sim$}}
        \raise1pt\hbox{$>$}}}

\begin{document}

\title{Sample variance in weak lensing: how many simulations are required?}

\author{Andrea Petri}
\email{apetri@phys.columbia.edu}
\affiliation{Department of Physics, Columbia University, New York, NY 10027, USA}
\affiliation{Physics Department, Brookhaven National Laboratory, Upton, NY 11973, USA}

\author{Zolt\'an Haiman}
\affiliation{Department of Astronomy, Columbia University, New York, NY 10027, USA}

\author{Morgan May}
\affiliation{Physics Department, Brookhaven National Laboratory, Upton, NY 11973, USA}

\date{\today}

\label{firstpage}

\begin{abstract}

Constraining cosmology using weak gravitational lensing consists of
comparing a measured feature vector of dimension $N_b$ with its
simulated counterpart. An accurate estimate of the $N_b\times N_b$
feature covariance matrix $\bb{C}$ is essential to obtain accurate
parameter confidence intervals. When $\bb{C}$ is measured from a set
of simulations, an important question is how large this set should
be. To answer this question, we construct different ensembles of $N_r$
realizations of the shear field, using a common randomization
procedure that recycles the outputs from a smaller number $N_s\leq
N_r$ of independent ray-tracing $N$--body simulations.  We study
parameter confidence intervals as a function of ($N_s,N_r$) in the
range $1\leq N_s\leq 200$ and $1\leq N_r\lesssim 10^5$.  Previous work
\citep{DodelsonSchneider13} has shown that Gaussian noise in the
feature vectors (from which the covariance is estimated) lead, at quadratic order, to an $O(1/N_r)$
degradation of the parameter confidence intervals. Using a variety of
lensing features measured in our simulations, including shear-shear
power spectra and peak counts, we show that cubic and quartic
covariance fluctuations lead to additional $O(1/N_r^2)$ error
degradation that is not negligible when $N_r$ is only a factor of few
larger than $N_b$. We study the large $N_r$ limit, and find that a
single, 240Mpc$/h$ sized $512^3$-particle $N$--body simulation ($N_s=1$) can be repeatedly recycled to
produce as many as $N_r={\rm few}\times10^4$ shear maps whose power spectra and high-significance peak counts can be treated as statistically independent. As a result, a small number of simulations ($N_s=1$ or $2$) is sufficient to forecast parameter confidence intervals at percent accuracy. 

\end{abstract}


\maketitle


\section{Introduction}
Weak gravitational lensing (WL) is a promising cosmological probe for
constraining the dark energy equation of state $w$, and has been
considered by a range of past (CFHTLens \citep{cfht1,cfht2}, COSMOS
\citep{cosmos}), ongoing (DES \citep{DES}) and future (LSST
\citep{LSST}, Euclid \citep{Euclid}, WFIRST \citep{WFIRST})
experiments. In an era where cosmology is data driven, accurate
numerical simulations of shear fields are becoming important for
several reasons, including assessing baryonic effects
\citep{BaryonXiuyuan,BaryonSemboloni,BaryonsWhite,BaryonsKnox,BaryonsZentner1,BaryonsZentner2},
the utility of non--Gaussian statistics
\citep{PeaksJan,MinkJan,MinkPetri,NG-Marian,NG-Jain1,NG-Jain2,NG-Jain3,NG-Refregier,NG-Dietrich}
and various systematic effects
\citep{MinkShirasaki,Sys-Bard,Sys-Chang,Sys-Huterer}.

A fundamental issue with predictions from simulations is that the
finite number of simulations naturally introduces fluctuations in the
forecasts, due to inevitable sample variance~\footnote{Sample variance is a broad term that has been used in the literature to describe a range of phenomena. Throughout this paper, we use it to refer to the fluctuations in an ensemble of simulations, which represent random realizations of the same initial conditions.}. In general,
quantities such as the mean or the variance of any feature (e.g. the
shear power spectrum at a multipole $\ell$), measured from a finite
set of simulations, will fluctuate, and can also suffer a bias.  While
biases in the estimates of both the mean and the variance have been
studied extensively, the impact of fluctuations in the variance has
received less attention.  These fluctuations have been shown to have non-negligible effects on estimates of
features covariances and hence on parameter constraints. In
particular, in the limit of Gaussian fluctuations, the parameter
confidence limits are degraged by a factor
$1+O(1/N_r)$~\citep{DodelsonSchneider13,Taylor12}.

This work studies these issues further, focusing on the number of
independent $N$--body simulations required for an accurate estimate of
the parameter constraints.  Ray-tracing simulations that resolve the
cosmic structures responsible for lensing on arcminute scales are
limited to physical sizes of hundreds of Mpc, and thus cover a solid
angle of only $O(10$ deg$^2)$.  As a result, many simulations are
required to tile a significant fraction of the sky, and to make
predictions for large ``all-sky'' surveys, such as the ones by DES,
LSST, Euclid, WFIRST.  In practice, this has led to the wide-spread
use of ``pseudo-independent'' realizations, i.e. a procedure in which
one randomizes and re-cycles the output of a single 3D simulation
multiple times.  In light of the forthcoming large surveys, it is
imperative to assess the statistical validity of this approach, and to
ask how many times a single simulation can be fairly recycled.  In
this paper, we address these questions with ensembles of up to
$N_r=10^5$ random realizations, extracted from up to $N_s=200$
independent ray-tracing $N$--body simulations. We focus in particular on
the parameter $w$, and on two different statistics: the (convergence)
power spectrum and the number counts of peaks.

This paper is organized as follows.  In \S~II, we summarize the shear
simulation methods we utilized, and describe the formalism we adopted
to forecast cosmological parameter constraints. We then vary the
number of simulations and the number of pseudo-independent
realizations, and present our main findings in \S~III. These results
are discussed further in \S~IV.  We offer our conclusions, and suggest
follow-up future work in \S~V.


\section{Methods}

\subsection{Ray-tracing simulations of the convergence field}
\label{shearsim}
In this section, we describe how we constructed our shear field
ensembles. Background galaxies at redshift $z_s$ are lensed by large
scale structures between $z=0$ and $z_s$. The shape distortions due to
the cosmic shear $\pmb{\gamma}$ can be computed in terms of the dark
matter gravitational potential $\Phi(\bb{x},z)$. Because the evolution
of $\Phi$ with redshift is non--linear, it needs to be computed with
numerical simulations. We make use of the public code \texttt{Gadget2}
\citep{Gadget2}, with which we run a sequence of $200$ independent
dark--matter--only $N$--body simulations that track the evolution of
the density fluctuations. We assume a standard $\Lambda$CDM background
universe with the parameters
$(\Omega_m,\Omega_\Lambda,h,w,\sigma_8,n_s)=(0.26,0.74,0.72,-1,0.8,0.96)$.
We fix the comoving size of the simulation box to $240\mathrm{Mpc}/h$,
and use $512^3$ particles, corresponding to a dark matter particle
mass of $\approx 10^{10}M_\odot$.

We assume a uniform galaxy distribution at a constant redshift $z_s=2$
(at which the simulation box has an angular size of $\theta_{\rm
  box}=3.5^\circ$) and we discretize the mass distribution
between $z_s$ and the observer at $z=0$ with a sequence of 46 two
dimensional lenses of thickness $80\mathrm{Mpc}/h$. The surface
density on each lens plane is computed by projecting the
three--dimensional density measured from \texttt{Gadget2}
snapshots. We then apply the multi--lens--plane algorithm (see
\citep{RayTracingHartlap,RayTracingJain} for example) to trace the
deflections of $n_{\rm ray}^2=2048^2$ light rays arranged on a square grid
of total size $\theta_{\rm box}$, from $z=0$ to $z_s$. This corresponds to a pixel angular resolution of $0.1^\prime$. 
Our implementation of this algorithm is part of the \texttt{LensTools}
computing package we have been developing \citep{LensTools}, and have
released under the \texttt{MIT} license. Many different realizations
$r$ of the same shear field $\pmb{\gamma}_r(\pmb{\theta})$ can be
generated by picking different lens planes that lie between the
observer and $z_s$. The randomization procedure we adopt is the
following (see \citep{Sato12} for reference):

\begin{itemize}
\item For each lens-plane redshift $z_l$, select the snapshot at $z_l$ from the $i$--th $N$--body simulation, where $i$ is a random integer $i\in [1,N_s]$.
\item Choose randomly between the three orthogonal directions ${\bb{n}_x,\bb{n}_y,\bb{n}_z}$: the lens plane will be perpendicular to this direction.
\item Choose the position of the plane along the snapshot: because the lens thickness is 1/3 the size of the box, we can cut three different slices of the simulation box for each orientation ${\bb{n}_x,\bb{n}_y,\bb{n}_z}$. This gives a total of 9 choices for generating a lens plane out of a single $N$--body snapshot.
\item Perform a periodic random shift of the lens plane along its two directions.
\item Repeat the above procedure for each lens-plane redshift $z_l$.
\end{itemize}  
This randomization procedure allows us to produce an (almost)
arbitrary number $N_r$ of shear realizations
$\pmb{\gamma}_r(\pmb{\theta})$. However, these realizations are not
guaranteed to be independent, if $N_s$ is not large enough. Using the
set of 200 independent $N$--body simulations, we construct different
ensembles with different choices of $N_s\in[1,200]$. Each of these
ensembles consists of the same number $N_r=1000$ of shear
realizations. We also build an additional ensemble with $N_s=1$ and
$N_r=10^5$ realizations. For each realization of each ensemble, we
reconstruct the convergence $\kappa_r(\pmb{\theta})$ from the trace of
the light-ray deflection Jacobian matrix, measured from the difference
in deflection angles between nearby light-rays
\citep{RayTracingHartlap,RayTracingJain,Sato12}.

We measure the $\kappa$ angular power spectrum $P^{\kappa\kappa}_r(\ell)$ defined as
\begin{equation}
\langle\tilde{\kappa}_r(\pmb{\ell})\tilde{\kappa}_r(\pmb{\ell}')\rangle = (2\pi)^2\delta_D(\pmb{\ell}+\pmb{\ell}')P^{\kappa\kappa}_r(\ell)
\end{equation}
As an additional summary statistic, we consider the counts of local
$\kappa$ maxima of a certain height $\kappa_0$, $n_r(\kappa_0)$
(hereafter \textit{peak counts}), with varying $\kappa_0$ chosen
between the minimum and maximum values measured from the maps
$(\kappa_{\rm min},\kappa_{\rm max})=(-0.06,0.45)$. Different choices
of $\kappa_0$ binning used in this work are outlined in Table
\ref{featuretable}.  The fact that the ensemble of $N_r$ realizations
is not completely independent if $N_s$ is not large enough can have an
effect on the covariance estimators of both $P^{\kappa\kappa}$ and
$n(\kappa_0)$.

To measure the cosmological dependence of the $\kappa$ peak counts, we
performed a set of additional ray--tracing simulations with different
combinations of the cosmological parameter triplet
$(\Omega_m,w,\sigma_8)$. A summary of the complete set of shear
ensembles used in this work is listed in Table \ref{simtable}.

\subsection{Cosmological parameter inference}
Let $\bbh{d}$ be a single estimate for a feature of dimension $N_b$,
$\bb{d}(\bb{p})$ be the true value of this feature at a point $\bb{p}$
in parameter space (which has a dimension $N_p$) and $\bb{C}$ be the
$N_b\times N_b$ feature covariance matrix. For the purpose of this
work $\bb{p}$ is the triplet $(\Omega_m,w,\sigma_8)$ and $\bb{d}$ is
one of the features -- either a power spectrum or a peak count
histogram -- in Table \ref{featuretable}. Although existing emulators
can be used, in principle, to compute both $\bb{d}(\bb{p})$ and
$\bb{C}$, the latter is more difficult, and typically only the mean,
$\bb{d}(\bb{p})$, has been computed to date
(refs.~\citep{coyote2,Nicaea}, but see an exception by
ref.~\citep{SchneiderKnoxCovariance}).  Estimating $\bb{C}$ from
simulations involves generating a series of mock realizations
$\bbh{d}_r$ with $r=1...N_r$ and computing the sample covariance
$\bbh{C}$,

\begin{equation}
\bb{\bar{d}} = \frac{1}{N_r}\sum_{r=1}^{N_r} \bbh{d}_r,
\end{equation}

\begin{equation}
\label{covest}
\bbh{C} = \frac{1}{N_r-1}\sum_{r=1}^{N_r} (\bbh{d}_r - \bar{\bb{d}}) (\bbh{d}_r - \bar{\bb{d}})^T.
\end{equation}
Assuming a normal feature likelihood, together with a flat prior on
the parameter space, the parameter posterior distribution
$\mathcal{L}(\bb{p}\vert\bbh{d}_{\rm obs})$ given an observed instance of
$\bbh{d}$, which we call $\bbh{d}_{\rm obs}$, follows from Bayes' theorem,
\begin{equation}
\label{posteriorbayes}
-2\log\mathcal{L}(\bb{p}\vert\bbh{d}_{\rm obs}) = [\bbh{d}_{\rm obs}-\bb{d}(\bb{p})]^T\bbh{C}^{-1}[\bbh{d}_{\rm obs}-\bb{d}(\bb{p})].
\end{equation}
For the sake of simplicity, we approximate the posterior as a Gaussian
around its maximum. This corresponds to Taylor-expanding the simulated
feature to first order around a point $\bb{p}_0$ (ideally the maximum
of Eq.~\ref{posteriorbayes}):
\begin{equation}
\bb{d}(\bb{p}) \approx \bb{d}_0 + \bb{d}_0^\prime(\bb{p}-\bb{p}_0).
\end{equation}
We chose $\bb{p}_0$ to be the triplet $(\Omega_m,w,\sigma_8)=(0.26,-1,0.8)$.
To measure the derivatives of the features $\bb{d}'_0$ with respect to
the cosmological parameters, we make use of the public code
\texttt{Nicaea} \citep{Nicaea} for the power spectrum, and we use an
independent simulation set (containing simulations with a variety of
different combinations of $(\Omega_m,w,\sigma_8)$, see Table
\ref{simtable}) for the peak counts.

\begin{table*}
\begin{center}
\begin{tabular}{c|c|c||c|c}
\toprule
$\Omega_m$ &  $w$ & $\sigma_8$ & $(N_s,N_r)$ & Number of $\kappa$ ensembles \\ \hline \hline
\midrule
0.26 & $-$1 & 0.8 & (1 to 200,1024) & 16 \\
0.26 & $-$1 & 0.8 & (1,128000) & 1 \\
0.29 & $-$1 & 0.8 & (1,1024) & 1 \\
0.26 & $-$0.8 & 0.8 & (1,1024) & 1 \\
0.26 & $-$1 & 0.6 & (1,1024) & 1 \\
\bottomrule
\end{tabular}
\end{center}
\caption{Summary of the shear ensembles used in this work. $N_s$ and
  $N_r$ refer to the number of independent $N$--body simulations, and the
  number of pseudo-independent realizations created from these
  simulations, respectively.}
\label{simtable}
\end{table*}

We can build the estimator for the posterior maximum $\bbh{p}$, given
the observation $\bbh{d}_{\rm obs}$, as follows:
\begin{equation}
\label{estimatormean}
\bbh{p} = \bb{p}_0 + \bbh{T}(\bbh{d}_{\rm obs}-\bb{d}_0),
\end{equation}

\begin{equation}
\bbh{T} = (\bb{d}_0'^T\bbh{C}^{-1}\bb{d}_0')^{-1}\bb{d}_0'^T\bbh{C}^{-1}.
\end{equation}
Because $\bbh{p}$ is estimated using a single noisy data instance
$\bbh{d}_{\rm obs}$, its estimate will be scattered around the true
value $\langle\bbh{p}\rangle_O$. In the following we use the
$\langle\rangle_O$ notation for expectation values taken with respect
to observations, while we keep the notation $\langle\rangle$ for
expectation values taken with respect to the simulations. Defining the
\textit{precision matrix} $\bbh{\Psi}=\bbh{C}^{-1}$, we can express
the estimator of the observational scatter in $\bbh{p}$:

\begin{equation}
\label{estimatorcovariance}
\h{\Sigma}_\bb{p} = \bbh{F}^{-1}\bb{d}_0'^T\bbh{\Psi}\langle(\bbh{d}_{\rm obs}-\bb{d}_0)(\bbh{d}_{\rm obs}-\bb{d}_0)^T\rangle_O\bbh{\Psi}\bb{d}_0'\bbh{F}^{-1},
\end{equation}

\begin{equation}
\label{estimatorfisher}
\bbh{F} = \bb{d}_0'^T\bbh{\Psi}\bb{d}_0'.
\end{equation}
Here we introduced the familiar Fisher matrix estimator $\bbh{F} =
\bb{d}_0'^T\bbh{\Psi}\bb{d}_0'$ and, for simplicity, we assumed
$\langle\bbh{d}_{\rm obs}\rangle_O=\bb{d}_0$, so that
$\langle(\bbh{d}_{\rm obs}-\bb{d}_0)(\bbh{d}_{\rm obs}-\bb{d}_0)^T\rangle_O=\bb{C}$.  When we perform an observation
$\bbh{d}_{\rm obs}$, the parameter estimate $\bbh{p}$ is a random draw
from a probability distribution with variance $\h{\Sigma}_\bb{p}$,
which inherits noise from the simulations. The noise in the covariance
estimator (Eq.~\ref{covest}) and in its inverse $\bbh{\Psi}$ propagate
all the way to the posterior (Eq.~\ref{posteriorbayes}), the parameter
estimate (Eq.~\ref{estimatormean}) and its variance
(Eq.~\ref{estimatorcovariance}). Following
refs.~\citep{DodelsonSchneider13,Taylor12,MasumotoWishart} we compute
the expectation value of Eq.~(\ref{estimatorcovariance}) over
simulations, $\langle\h{\Sigma}_\bb{p}\rangle$, up to $O(1/N_r^2)$ by
expanding Eq.~(\ref{estimatorcovariance}) to quartic order in the
statistical fluctuations of $\bbh{\Psi}$. Denoting the true
parameter covariance, i.e. the usual inverse Fisher matrix, as
$\Sigma_\bb{p}=\bb{F}^{-1}$, we find the result~\footnote{The details
  of the calculation are given in Appendix A.}:

\begin{widetext}
\begin{equation}
\label{quarticdegradation}
\langle\h{\Sigma}_\bb{p}\rangle = \Sigma_\bb{p}\left[1+\frac{N_b-N_p}{N_r}+\frac{(N_b-N_p)(N_b-N_p+2)}{N_r^2}\right] + O\left(\frac{1}{N_r^3}\right).
\end{equation}
\end{widetext}
Although we truncated the expansion to second order in $1/N_r$, an exact expression for $\langle\h{\Sigma}_\bb{p}\rangle$ has been proposed by \citep{Taylor14}
\begin{equation}
\label{empiricalexact}
\langle\h{\Sigma}_\bb{p}\rangle_{\rm empirical} = \Sigma_\bb{p}\left(\frac{N_r-2}{N_r-N_b+N_p-2}\right)
\end{equation}
This empirical expression reduces to equation (\ref{quarticdegradation}) when expanded at order $O(1/N_r^2)$ but, to our knowledge, no first principles proof of its correctness exists. 
Next, we restrict ourselves to the large $N_r$ limit, and we further
investigate the behavior of the $O(1/N_r)$ term. We consider three
cases \footnote{The details of these calculations are given in Appendix B}:

\begin{enumerate}
\item If the true data covariance $\bb{C}$ is known, the estimator in
  eq.~(\ref{estimatorcovariance}) is biased, and the dominant
  contribution of the bias comes from the second order fluctuations in
  $\bbh{\Psi}$. Once the expectation values over simulations are
  taken, the bias sums up to

\begin{equation}
\label{dodelsonscaling}
\langle\h{\Sigma}_\bb{p}\rangle=\Sigma_\bb{p}\left(1+\frac{N_b-N_p}{N_r}\right).
\end{equation}
This is the result obtained by ref.~\citep{DodelsonSchneider13}.

\item Usually the true data covariance is unknown, and it is tempting
  to plug in its estimator $\bbh{C}$, measured from the same
  simulation set we use to compute $\bbh{\Psi}$. This approach has
  been used before in the literature
  (e.g.~\citep{MinkPetri,MinkShirasaki}). If this is done without
  correcting for the bias in $\bbh{\Psi}$ (see Ref.~\citep{Taylor12}
  and eq.~\ref{firstmoment} below), the parameter variance will have a
  contribution from both the second and first-order fluctuations in
  $\bbh{\Psi}$, which now have a nonzero expectation value. In this
  case the bias sums up to
\begin{equation}
\label{mockscalinguncorrected}
\langle\h{\Sigma}_\bb{p}\rangle=\Sigma_\bb{p}\left(1-\frac{N_b-N_p}{N_r}\right).
\end{equation}

\item If we repeat the same exercise as above, but we correct for the
  bias in the precision matrix estimator, we are left with

\begin{equation}
\label{mockscalingcorrected}
\langle\h{\Sigma}_\bb{p}\rangle=\Sigma_\bb{p}\left(1+\frac{1+N_p}{N_r}\right).
\end{equation}

\end{enumerate} 
The error degradation in each parameter $p$, at leading order, scales
as $D/N_r$, where $D=N_b-N_p$, $N_p-N_b$, and $1+N_p$ for cases 1, 2,
and 3, respectively.  Note that in the last case, which is most
relevant when fitting actual data, the estimated degradation turns out
to be too optimistic: the parameter estimate $\bbh{p}$ has a variance
whose noise grows linearly with $N_b$ (eq.~\ref{quarticdegradation}),
whereas the degradation estimated via eq. (\ref{mockscalingcorrected})
is constant with $N_b$. This can lead to underestimation of error
bars, which can be mistakenly interpreted as a parameter bias. We test
scaling relations of the form
\begin{equation}
\label{ourscaling}
\langle\h{\sigma}_p^2\rangle = \sigma^2_{p,\infty}(N_s)\left(1+\frac{D}{N_r}\right)
\end{equation}
against our simulations, in the limits of both high and low $N_r$.  We
indicate the diagonal elements of $\h{\Sigma}_{\bb{p}}$ as
$\h{\sigma}^2_p=\mathrm{diag}(\h{\Sigma}_{\bb{p}})$ and we indicate by
$\sigma^2_{p,\infty}$ the expectation value of the variance of each
parameter in the limit of an infinite number of realizations
$N_r\rightarrow\infty$.  We call $D$ the \textit{effective
  dimensionality} of the feature space (which, as seen before, can be
negative in some pathological cases).  We compute the expectation
values of $\h{\sigma}^2_p$ (eqs.~\ref{estimatorcovariance} and
\ref{ourscaling}) by averaging over 100 random resamplings of our
shear ensembles. For the true feature covariance matrix
$\langle(\bbh{d}-\bb{d}_0)(\bbh{d}-\bb{d}_0)^T\rangle=\bb{C}$ we use
the estimated covariance from a grand ensemble built with the union of
all the ensembles with different $N_s$.

The true parameter variance $\sigma^2_{p,\infty}(N_s)$ in principle
can depend on the number of independent $N$--body simulations $N_s$,
which appears in the randomization procedure described in
\S~\ref{shearsim} above. This is because if $N_s$ is not large enough,
the different shear realizations cannot be all independent, and hence
the true variance $\sigma^2_{p,\infty}(N_s\rightarrow\infty)$ cannot
be recovered for low $N_s$ even if $N_r$ is arbitrarily large. In the
next section, we present our main findings.


\section{Results} 


\begin{figure*}
\includegraphics[scale=0.4]{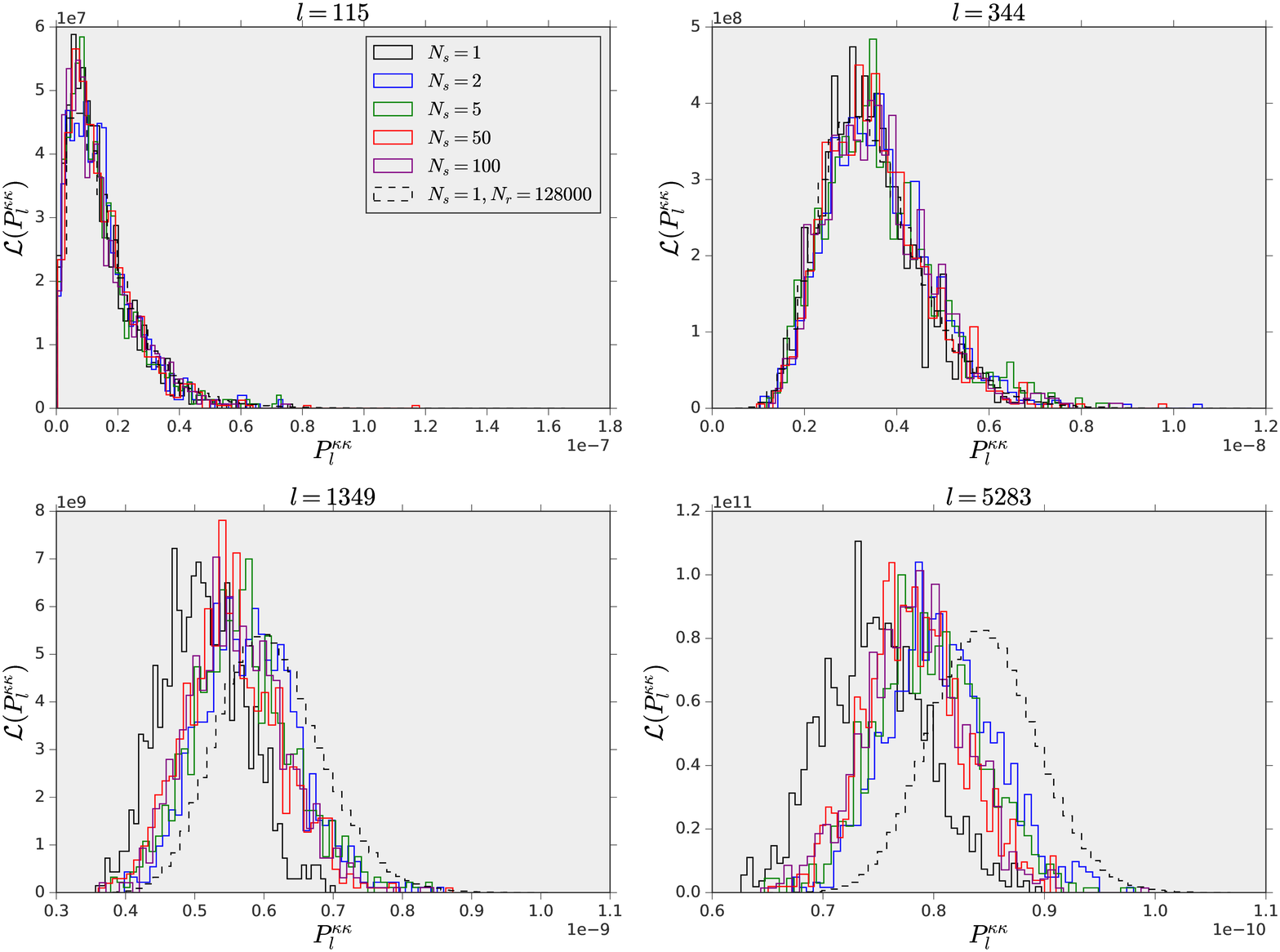}
\caption{PDF of the $\kappa$ power spectrum
  $\mathcal{L}(P_l^{\kappa\kappa})$ at four selected multipoles
  $\ell=115,344,1349,5283$, for different shear ensembles constructed
  from on $N_s=$1 (black), 2 (blue), 5 (green), 50 (red), and 100
  (purple) independent $N$--body simulations. Each curve is based on
  $N_r=1024$ realizations.  The dashed black curves correspond to
  ensembles generated with $N_s=1$ and $N_r=128000$. For $N_s\geq2$,
  the distributions appear similar to the eye; this similarity is
  confirmed by the comparisons in Figures \ref{means_ns} and
  \ref{ps_var} below.}
\label{ps_pdf}
\end{figure*}

\begin{figure}
\includegraphics[scale=0.3]{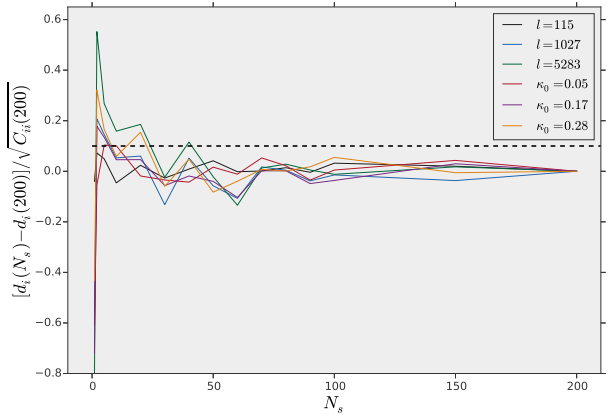}
\caption{The mean value $\bb{\bar{d}}$ of various features, measured
  from ensembles created from different numbers $N_s$ of simulations.
  For each case, the difference compared to the mean in the $N_s=200$
  ensemble is shown, in units of the statistical error measured in the
  $N_s=200$ ensemble.  The colored curves refer to shear--shear power
  spectra measured at $\ell=115$ (black), 1027 (cyan), and 5283
  (green), and peak counts with heights $\kappa_0=0.05$ (red), 0.17
  (purple), and 0.28 (orange). The $\kappa$ bin width for the peak
  counts has been fixed to $\Delta\kappa=0.011$. The dashed black line
  shows a level of $0.1\sigma$ accuracy for reference.  For $N_s\geq
  2$, the means are statistically indistinguishable (even at
  $\sim0.1\sigma$) from those in the ensemble with $N_s=200$.}
\label{means_ns}
\end{figure}

\begin{figure}
\includegraphics[scale=0.3]{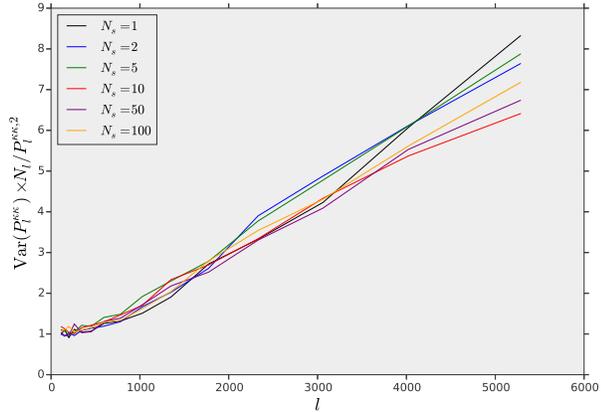}
\caption{Variance of the $\kappa$ power spectrum as a function of the
  multipole $\ell$, in units of the expected Gaussian variance from
  equation (\ref{gaussianvar}). The variance is measured from
  different shear ensembles based on $N_s=$1 (black), 2 (blue), 5
  (green), 10 (red), 50 (purple), or 100 (orange) N--body simulations.
  Non--Gaussianities of the underlying structures increase the
  variance on small scales, but no clear trend with $N_s$ can be
  identied on any scale.}
\label{ps_var}
\end{figure}

\begin{figure}
\includegraphics[scale=0.3]{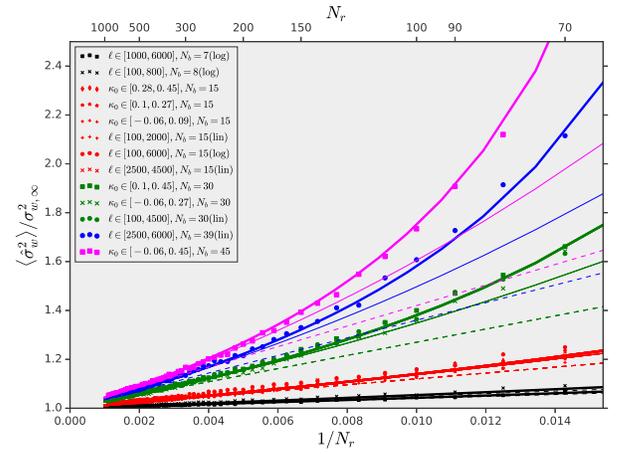}
\caption{Expectation value of the variance of $w$ computed from
  equation (\ref{estimatorcovariance}), shown as a function of
  $1/N_r$. The different symbols and colors correspond to the features
  listed in Table~\ref{featuretable}.  The dashed and this solid curves
  show the analytic predictions from equation
  (\ref{quarticdegradation}) at orders $O(1/N_r)$ and $O(1/N_r^2)$,
  respectively. The thick solid curves show the empirical predictions from equation (\ref{empiricalexact}). 
  The asymptotic variance $\sigma^2_{w,\infty}$ has been
  computed from a linear regression of $\langle\h{\sigma}^2_w\rangle$
  vs $1/N_r$ for $N_r>500$. The figure clearly shows that terms beyond
  $O(1/N_r)$ need to be considered, unless $N_r\gg N_b$.}
\label{curvingnb}
\end{figure}

\begin{figure}
\includegraphics[scale=0.3]{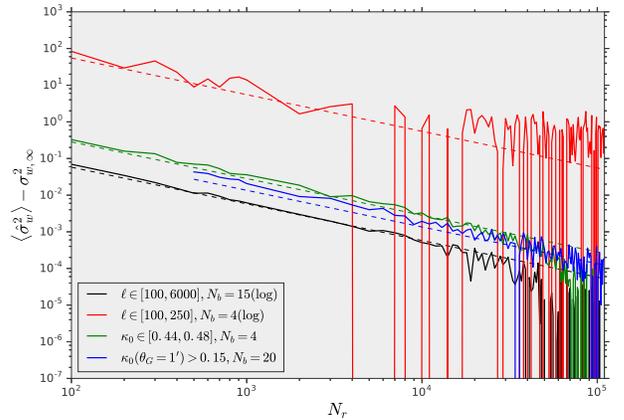}
\caption{Bias in the variance of $w$,
  $\langle\h{\sigma}^2_w\rangle-\sigma^2_{w,\infty}$, as a function of
  the number of realizations $N_r$ used to estimate the covariance
  (eq.~\ref{covest}). The figure shows both the trend measured in the
  simulations (solid lines) and their scaling expected from
  Eq.~(\ref{ourscaling}) with $D=N_b-N_p$ (dashed line). The
  asymptotic variance $\sigma^2_{w,\infty}$ has been estimated to be the value $\langle\h{\sigma}^2_w\rangle(N_r=10^5)$. Different features are considered: power spectra with logarithmically spaced $\ell\in[100,6000]$ (black), $\ell\in[100,250]$ (red), peak counts in the unsmoothed maps with height $\kappa_0\in[0.44,0.48]$ (green) and peak counts in the smoothed maps (with a Gaussian kernel of size $\theta_G=1^\prime$) with height $\kappa_0>0.15$ (blue). No deviations from the expected $1/N_r$ behavior are observed up to $N_r\approx{\rm few}\times10^4$, except for the large-scale power spectrum, in which case the deviations occur much earlier ($N_r\approx 10^3$).}
\label{wvar_nr}
\end{figure}

\begin{figure}
\includegraphics[scale=0.3]{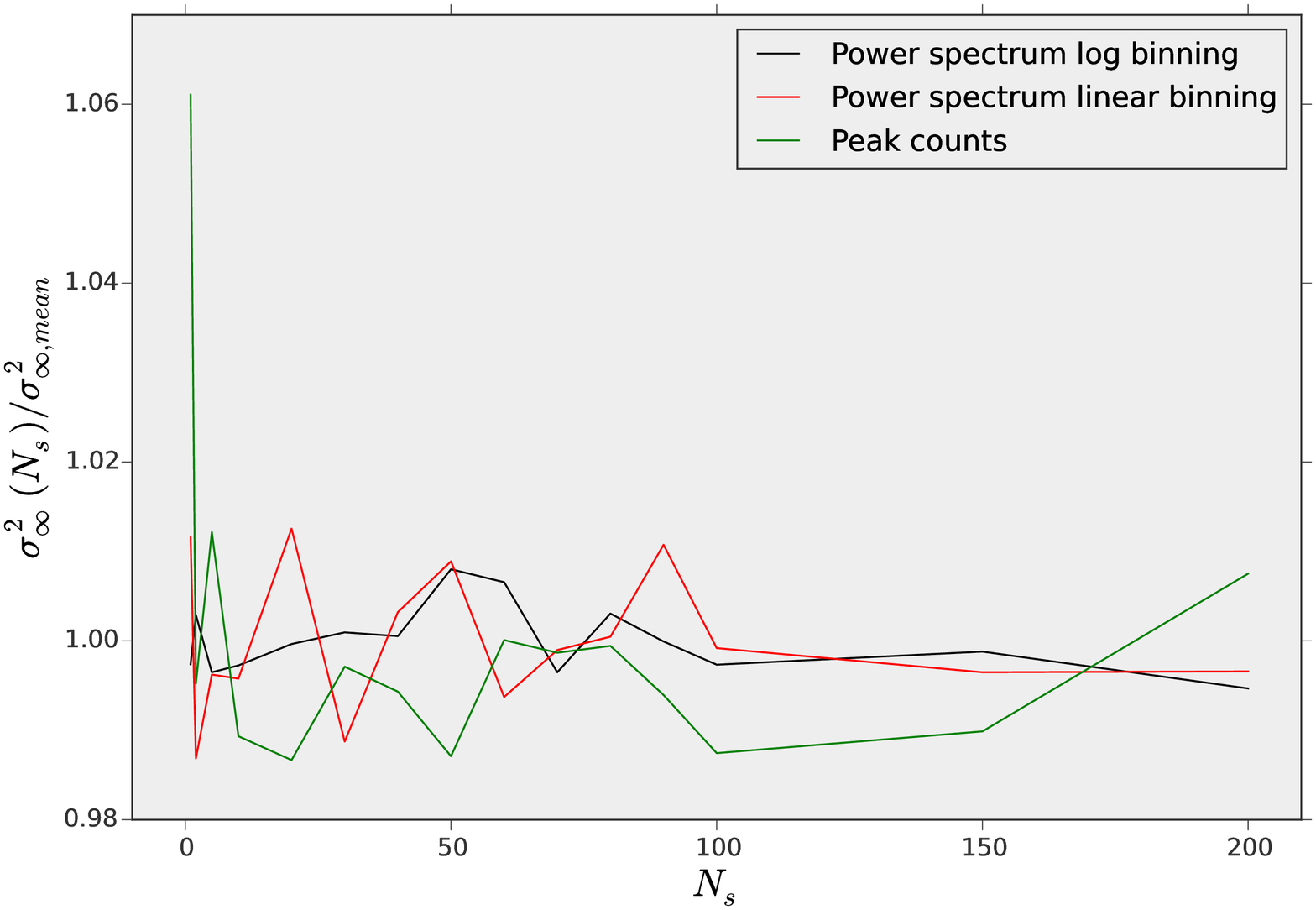}
\caption{The variance of $w$ in the limit of $N_r\rightarrow\infty$
  (measured from the intercept of the fit $\sigma^2_w$ vs $1/N_r$),
  varying the number of simulations $N_s$ used in the ensemble to
  estimate the covariance (eq.~\ref{covest}). We show the dependence
  of $\sigma_{w,\infty}^2(N_s)$ in units of the mean over the union of 16
  ensembles with different $N_s$ and $N_r=1024$, for the power spectrum
  logarithmically binned (black, $N_b=15,\ell\in[100,6000]$), the
  power spectrum linearly binned (red,$N_b=39,\ell\in[100,6000]$) and
  the peak counts (green,$N_b=45,\kappa_0\in[-0.06,0.45]$).  No trend
  with $N_s$ is seen for $N_s\geq 2$, and the differences are only of
  order 1\%.}
\label{wvar_ns}
\end{figure}

\begin{table*}
\begin{center}
\begin{tabular}{ccccc}
\toprule
\textbf{Feature} &  \textbf{Specifications} & $N_b$ &  \textbf{Symbol} & \textbf{Color} \\ \hline \hline
\midrule
Power Spectrum, log binning  & $\ell \in [100,800] $ & 8 & $\times$ & black  \\ 
Power Spectrum, log binning  & $\ell \in [1000,6000] $ & 7 & $\blacksquare$ & black  \\ 
Power Spectrum, log binning  & $\ell \in [100,6000] $ & 15 & \textcolor{red}{$\bullet$} & \textcolor{red}{red}  \\
Power Spectrum, linear binning  & $\ell \in [100,2000] $ & 15 & \textcolor{red}{$+$} & \textcolor{red}{red}  \\ 
Power Spectrum, linear binning  & $\ell \in [2500,4500] $ & 15 & \textcolor{red}{$\times$} & \textcolor{red}{red}  \\
Power Spectrum, linear binning  & $\ell \in [100,4500] $ & 30 & \textcolor{OliveGreen}{$\bullet$} & \textcolor{OliveGreen}{green}  \\ 
Power Spectrum, linear binning  & $\ell \in [100,6000] $ & 39 & \textcolor{blue}{$\bullet$} & \textcolor{blue}{blue}  \\ \hline
Low peaks  & $\kappa_0 \in [-0.06,0.09] $ & 15 & \textcolor{red}{$+$} & \textcolor{red}{red}  \\ 
Intermediate peaks  & $\kappa_0 \in [0.1,0.27] $ & 15 & \textcolor{red}{$\bigstar$} & \textcolor{red}{red}  \\ 
High peaks  & $\kappa_0 \in [0.28,0.45] $ & 15 & \textcolor{red}{$\diamond$} & \textcolor{red}{red}  \\
Low+Intermediate peaks  & $\kappa_0 \in [-0.06,0.27] $ & 30 & \textcolor{OliveGreen}{$\times$} & \textcolor{OliveGreen}{green}  \\
Intermediate+High peaks  & $\kappa_0 \in [0.1,0.45] $ & 30 & \textcolor{OliveGreen}{$\blacksquare$} & \textcolor{OliveGreen}{green}  \\
All peaks  & $\kappa_0 \in [-0.06,0.45] $ & 45 & \textcolor{magenta}{$\blacksquare$} & \textcolor{magenta}{magenta}  \\ \hline
\bottomrule
\end{tabular}
\end{center}
\caption{Catalog of feature types used in this work, along with the
  chosen number of bands $N_b$ and the plot legends for Figure
  \ref{curvingnb}.}
\label{featuretable}
\end{table*}


In this section we present the main results of this work. We show the
qualitative behavior of a variety of feature $\bbh{d}_r$ probability
distribution functions (PDFs) in ensembles built with different $N_s$
and $N_r$. In Figure~\ref{ps_pdf}, we show the PDF of the power
spectrum at four selected multipoles, spanning the linear ($\ell=115$)
to the nonlinear ($\ell=5283$) regime. In Figure~\ref{means_ns}, we
shows the ensemble mean for these power spectra, as well as for peak
counts of three different $\kappa_0$ heights (corresponding to
$\approx 2-13\sigma$ peaks), as a function of $N_s$.  In
Figure~\ref{ps_var}, we show the variance of the power spectrum at
each multipole, as a function of $N_s$, in units of the variance
expected if the convergence $\kappa$ was a Gaussian random field
\begin{equation}
\label{gaussianvar}
\mathrm{Var}(P^{\kappa\kappa}_\ell) = \frac{(P^{\kappa\kappa}_\ell)^2}{N_{\rm eff}(\ell).}
\end{equation}
Here $N_{\rm eff}(l)$ is the number of independent modes used to
estimate the power spectrum at $\ell$.  

In practice, we measure $P^{\kappa\kappa}_\ell$ on the Fourier
transform of the pixelized simulated map $\kappa_r(\pmb{\theta})$,
using the FFT algorithm, and some care must be taken to count the
number of modes $N_{\rm eff}(\ell)$ correctly. Each pixel $(i_x,j_y)$
in Fourier space corresponds to a mode
$(\ell_x,\ell_y)=2\pi(i_x,i_y)/\theta_{\rm box}$, with $i_x=-n_{\rm
  ray}/2,...,n_{\rm ray}/2$ and $i_y=0,...,n_{\rm ray}/2$.  Here
$n_{\rm ray}=2048$ is the linear number of pixels on the ray--traced
convergence maps.  We count the number of pixels $N(\ell)$ that fall
inside a multipole bin $(\ell_1,\ell_2)$. Because the $\kappa$ field
is real, the modes $(\pm \ell_x,0)$ are not independent. If we let
$N(\ell,\ell_y=0)$ be the number of non--independent modes, the
effective number of independent modes for the variance is given by

\begin{equation}
N_{\rm eff}(\ell) = \frac{N^2(\ell)}{N(\ell)+N(\ell,\ell_y=0)}.
\end{equation}
This correction is important at low $\ell$, where pixelization effects
are non-negligible; $N_{\rm eff}(\ell\gg2\pi/\theta_{\rm box})\approx
N(\ell)$.

In Figures~\ref{curvingnb} and \ref{wvar_nr}, we show the dependence
of the confidence range $\langle\h{\sigma}^2_w\rangle$ on $N_r$,
derived from the features used in this work (see Table
\ref{featuretable} for a comprehensive list). 
Figure~\ref{curvingnb} shows the
behavior in the limit of a large number $N_r\gg500$ of realizations,
and compares it with the scaling of the form in
equation~(\ref{ourscaling}).
Figure \ref{wvar_nr} shows the large $N_r$ trends of the $w$ constraint.
Figure~\ref{curvingnb} illustrates the behavior at relatively low
$N_r$, and compares $\langle\h{\sigma}^2_w\rangle$ measured directly
from the simulations with the analytic expectations from
equation~(\ref{quarticdegradation}). Finally, in Figure~\ref{wvar_ns}, we show
how the $w$ confidence limit changes with $N_s$.


\section{Discussion}

In this section we discuss our main findings and their
implications. Figure \ref{ps_pdf} shows that, although different
choices of $N_s$ do not affect the power spectrum PDF on large scales
(top two panels), there are some qualitative differences on smaller
scales (bottom two panels). On these smaller scales, shear ensembles
built from $N_s=1$ do not produce the same statistical behavior as
ensembles built with larger $N_s$. In particular, looking at the black
curves, we see that the $N_s=1$ ensembles exibit large shifts with
respect to the other PDFs to lower power, including the locations of
the peaks of the PDFs.  We attribute these offsets to large (random)
statistical errors.

Interestingly, we need as few as $N_s=2$ simulations to recover the
right PDF for the small--scale power spectrum.  Figure~\ref{means_ns}
shows that multiple independent $N$--body simulations $N_s\geq2$ are
indeed necessary for measuring the means of feature ensembles to an
accuracy corresponding to 10\% of the statistical error. The number of
required simulations $N_s$ depends on the feature type and ranges from
a few ($N_s=1$ or 2) for the power spectrum at low multipoles
($\ell\lesssim 500$) to $N_s\approx 30-50$ for the power spectrum at
larger multipoles ($\ell\gtrsim1000$) or peak counts above a high
threshold ($\kappa_0\approx0.3$).  On the other hand, relaxing the
required accuracy to 50\% of the statistical error, we find $N_s=2$ to
be always sufficient. As pointed out by \citep{NbodySample}, the box size used for the $N$--body simulations can also play an important role in the accuracy of the power spectrum ensemble means.

Figure \ref{ps_var} shows the variance of convergence power spectrum
computed from different ensembles, in units of the Gaussian
expectation. We find that, even with $N_s=1$, we are able to recover
the known result that non--Gaussian structures increase the variance
significantly on small scales (see \citep{Sato12,TakadaSpergel14} for
reference).  Our results are in fact in excellent quantitative
agreement with \citep{Sato12}, which used $N_s=400$ independent $N$--body
simulations.  This result is highly encouraging, suggesting that
individual $N$--body runs can be recycled repeatedly.  However, it is
not sufficient by itself to conclude that $N_s$ does not impact the
parameter inferences, since these depend on the cross band
covariances.

Figure \ref{curvingnb} investigates the parameter errors. This figure
shows that error degradation estimates truncated at order $O(1/N_r)$
are too optimistic when the number of simulations $N_r$ used to
measure the covariance is only a factor of few larger than the
dimension of the feature space $N_b$. In these cases effects coming
the next--to--leading orders $O(1/N_r^2)$ become non--negligible on
constraint degradation. In particular, we find that already for
$N_b=30$ and $N_r\sim100$, the error degradation estimates to the next
leading order, $O(1/N_r^2)$, remain too optimistic. Accurate analytic
estimates in this regime would require at least terms of order
$O(1/N_r^3)$, which come from higher--than--quartic $\bbh{\Psi}$
fluctations.

In Figure \ref{wvar_nr}, we examine how the degradation in the $w$
constraint depends on the number of simulations used to estimate the
covariance, in the limit of large $N_r$. We find excellent agreement
with the expected scaling (eq.~\ref{ourscaling}) up to $N_r\sim {\rm few}\times10^4$ when using the $\kappa$ power spectrum in the multipole range $\ell\in[100,6000]$. The same behavior is observed when considering the high-significance peak counts ($>10\sigma$ for unsmoothed maps and $>5\sigma$ for $1^\prime$ smoothed maps). As the figure shows, around these values of $N_r$ the $\langle\hat{\sigma}^2_w\rangle - \sigma^2_{w,\infty}$ curve
becomes noisy and reaches negative values. This is a clear indication that the $1/N_r$ behavior is broken and a plateau in
$\langle\hat{\sigma}^2_w\rangle$ is reached. The negative values in
the plot are a consequence of the noise in the estimation of this
plateau value (or equivalently in the estimated value of
$\sigma^2_{w,\infty}$). 

We conclude that a single $N$--body
simulation is sufficient to construct an ensemble of up to a
few$\times 10^4$ mutually independent convergence power spectra. For
$N_r\gg10^4$, the shear realizations can no longer be considered
independent. We emphasize that the precise value of this $N_r$ will depend on the size of the simulation box (which, in our case, is (240Mpc$/h$)$^3$, with $512^3$ particles) and also on the range of multipoles $\ell$ used to constrain the parameters. Figure \ref{wvar_nr} shows that when we infer $w$ only from large--scale modes, $\ell\lesssim250$, the plateau is reached at least an order of magnitude earlier in the number of realizations. In other words, the number of independent power spectra we can generate decreases as we increase the spatial scales of interest. This is due to the fact that, because of the finite box size, the number of independent lens plane shifts (as described in \S~\ref{shearsim}) decreases as the mode size approaches the size of the box.    Similarly, one may expect that the independence in the statistics of high-amplitude peaks, which are predominantly produced by single massive halos, may be compromised by these halos being present repeatedly, in many of the pseudo-independent realizations.  However, Figure \ref{wvar_nr} shows that this is not the case: the peak count statistics are shown at $\kappa$ thresholds corresponding to massive ($\approx 10^{15}~{\rm M_\odot}$) halos, yet there is no evidence that the independence of the maps breaks down until $N_r=$few$\times 10^4$.  Apparently, randomly projected structures, which vary from realization-to-realization, contribute significantly to the statistics of these high peaks.

Figure \ref{wvar_ns} shows how the ``true'' $w$ constraint (in the
limit $N_r\rightarrow \infty$; or equivalently the $w$ constraint with
the known $N_r$ dependence factored out), depends on $N_s$. We find
that, in the range $N_s\in[1,200]$ the inferred $w$--variance
$\sigma_{w,\infty}^2$ fluctuates stochastically only by 1\%, and does
not show any trend with $N_s$.

Finally, we found that when we estimate the data covariance $\bb{C}$
from the same simulation set used to measure $\bbh{\Psi}$, the
effective dimensionality $D$ decreases with increasing $N_b$ in the
case where the $\bbh{\Psi}$ bias is not corrected
(eq.~\ref{mockscalinguncorrected}).  This $N_b$-dependence disappears
when the bias is corrected (eq.~\ref{mockscalingcorrected}).  This
fact that should be taken into consideration when forecasting
parameter errors purely from simulations, as the errors will otherwise
be underestimated. A similar conclusion was reached by
\citep{Hartlap07} (although their paper did not address the impact of
using the same simulation set for $\bb{C}$ and $\bbh{\Psi}$).


\section{Conclusions}

In this work, we have examined the effect of forecasting cosmological
constraints based on shear ensembles generated from a finite number of
$N$--body simulations.  Our main results can be summarized as follows:
\vspace{0.4\baselineskip}
\begin{itemize}

\item When the feature covariance matrix is measured from simulations,
  parameter constraints are degraded. This degradation is appreciably
  larger than the $O(1/N_r)$ computed by \citep{DodelsonSchneider13}
  when the number of realizations $N_r$ is only a factor of few larger
  than the feature vector size $N_b$.
\item We can recycle a single 240Mpc$/h$ $N$--body simulation to produce an
  ensemble of $O(10^4)$ shear maps whose small-scale power spectra and high-signficiance peak counts are statistically independent. The mean feature measured from a shear ensemble, though, could be inaccurate if only one $N$--body simulation is used.
\item As few as one or two independent $N$--body simulations are
  sufficient to forecast $w$ error bars to 1\% accuracy, provided that
  a sufficiently large number $N_r$ of realizations are used to
  measure feature covariances.  In particular, provided that biases
  in the inverse covariance are corrected, percent--level forecasts
  require $N_r\gsim 100 (N_b-N_p)$ realizations.
\item Depending on the feature type used to constrain cosmology, a
  larger number of $N$--body simulations might be needed to measure
  accurate ensemble means to an accuracy corresponding 10\% of the
  statistical error. If this accuracy requirement is relaxed to 50\%
  of the statistical error, we find that as low as $N_s=2$ simulations
  are sufficient for the feature types we consider in this work.
\end{itemize}
Future extensions of this work should involve extending our analysis
to a larger set of cosmological parameters, and to more general
feature spaces, such as the ones that characterize non--Gaussian
statistics (e.g. including higher moments of the $\kappa$ field,
Minkowski Functionals, and higher-order $\kappa$ correlators). While
our results are highly encouraging, and suggest that a single $N$--body
simulation can be recycled repeatedly, to produce as many as $10^4$
independent shear power spectra or peak count histograms. In order to
scale our results to large future surveys, such as LSST, it will be
necessary to determine if our findings hold when challenged by larger
and higher-resolution $N$--body simulations \citep{Qcontinuum}.


\section*{Acknowledgements}
We thank Lam Hui for useful discussions.  The simulations in this work
were performed at the NSF XSEDE facility, supported by grant number
ACI-1053575, and at the New York Center for Computational Sciences, a
cooperative effort between Brookhaven National Laboratory and Stony
Brook University, supported in part by the State of New York. This
work was supported in part by the U.S. Department of Energy under
Contract Nos. DE-AC02-98CH10886 and DE-SC0012704, and by the NSF Grant
No. AST-1210877 (to Z.H.) and by the Research Opportunities and
Approaches to Data Science (ROADS) program at the Institute for Data
Sciences and Engineering at Columbia University (to Z.H.).  

\bibliography{ref}

\section*{Appendix A: cubic and quartic covariance fluctuations}
\label{appendixA}

The goal of this appendix is to give a derivation of
eq. (\ref{quarticdegradation}). When the simulated feature vector
$\bbh{d}_r$ is drawn from a Gaussian distribution, the covariance
estimator $\bbh{C}$ follows the Wishart distribution, and its inverse
$\bbh{\Psi}$ follows the inverse Wishart distribution (see
Ref.~\citep{Taylor12} for analytical expressions for these probability
distributions). Computing expectation values of
eq. (\ref{estimatorcovariance}) over the inverse Wishart distribution
is not possible analytically, and a perturbative expansion is
necessary. Writing $\bbh{\Psi}=\bb{\Psi}+\delta\bbh{\Psi}$, we can
expand eq.~(\ref{estimatorcovariance}) in powers of
$\delta\bbh{\Psi}$. The expectation value of each term in this
expansion can be calculated in terms of moments of the inverse Wishart
distribution. Ref.~\citep{MasumotoWishart} provides a general
framework to compute these moments, and give exact expressions for
moments up to quartic order. First, let us expand the inverse of the
Fisher matrix estimator (eq.~\ref{estimatorfisher}) in powers of
$\delta\bbh{\Psi}$. The $n$--th order of this expansion will be

\begin{equation}
\label{nthterm}
\delta\bbh{F}^{-1}_{(n)} = (-1)^{n}(\bb{F}^{-1}\delta\bbh{F})^n\bb{F}^{-1}
\end{equation}
with
\begin{equation}
\delta\bbh{F} = \bb{d}_0'^T\delta\bbh{\Psi}\bb{d}_0'
\end{equation}
Using eq. (\ref{nthterm}), we can expand
eq. (\ref{estimatorcovariance}) to an arbitrary order in
$\delta\bbh{\Psi}$, take the expectation values of the fluctuations
over the inverse Wishart distribution, and finally arrive at
eq.~(\ref{quarticdegradation}). We use the notation $\nu\equiv N_r-1$
and $\gamma\equiv (\nu-N_b-1)/2$, and we indicate with capital letters
pairs of matrix indices, for example $I=(i_1,i_2)$, where
$i_a=1..N_b$. The main results we utilize from
ref.~\citep{MasumotoWishart} regarding the first four moments are (up
to order $O(1/\nu^2)$)

\begin{widetext}

\begin{equation}
\label{firstmoment}
\langle\h{\Psi}_I\rangle = \frac{\nu}{2\gamma}\Psi_I
\end{equation}

\begin{equation}
\label{secondmoment}
\langle\delta\h{\Psi}_I\delta\h{\Psi}_J\rangle = \frac{\nu^2\Psi_I\Psi_J + \nu^2\gamma\Psi_{\{I}\Psi_{J\}}}{4\gamma^2(\gamma-1)(2\gamma+1)}
\end{equation}

\begin{equation}
\label{thirdmoment}
\langle\delta\h{\Psi}_I\delta\h{\Psi}_J\delta\h{\Psi}_K\rangle = \frac{\nu^3\Psi_{\{I}\Psi_J\Psi_{K\}}}{8\gamma(\gamma-1)(\gamma-2)(\gamma+1)(2\gamma+1)}
\end{equation}

\begin{equation}
\label{fourthmoment}
\langle\delta\h{\Psi}_I\delta\h{\Psi}_J\delta\h{\Psi}_K\delta\h{\Psi}_L\rangle = \frac{\nu^4(2\gamma^2-5\gamma+9)\Psi_{\{I}\Psi_{J\}}\Psi_{\{K}\Psi_{L\}}}{16\gamma(\gamma-1)(\gamma-2)(\gamma-3)(2\gamma-1)(\gamma+1)(2\gamma+1)(2\gamma+3)}.
\end{equation}

\end{widetext}
Here the curly bracket notation is a shorthand for a symmetrization over pair of indices: for example
\begin{equation}
\Psi_{\{I}\Psi_{J\}} = \Psi_{i_1j_1}\Psi_{i_2j_2} + \Psi_{i_1j_2}\Psi_{i_2j_1}
\end{equation}
Eq. (\ref{firstmoment}) expresses the bias in the $\bbh{\Psi}$
estimator that already appears in the literature~\citep{Hartlap07}. If
we want to use the bias-corrected $\bbh{\Psi}$ estimator (required for
the perturbative expansion of eq.~\ref{estimatorcovariance}), we need
to apply an additional factor of $(2\gamma/\nu)^n$ to
eqs. (\ref{firstmoment}--\ref{fourthmoment}), where $n$ is the order
of the moment up to which we are applying the correction. If we limit
ourselves to computing the expectation value of
eq. (\ref{estimatorcovariance}) up to order $O(1/\nu^2)$, we do not
need to worry about this correction for
eqs. (\ref{thirdmoment}--\ref{fourthmoment}), as the dominant term
here is already $O(1/\nu^2)$. The next step is expanding
eq. (\ref{estimatorcovariance}) in powers of $\delta\bbh{\Psi}$ up to
fourth order: this is easily done:

\begin{widetext}
\begin{equation}
\label{fullpertexpansion}
\h{\Sigma}_\bb{p} = \left(\bb{F}^{-1}+\sum_{n=1}^4\delta\bbh{F}^{-1}_{(n)}\right)\bb{d}'^T_0(\bb{\Psi}+\delta\bbh{\Psi})\bb{C}(\bb{\Psi}+\delta\bbh{\Psi})\bb{d}'_0\left(\bb{F}^{-1}+\sum_{n=1}^4\delta\bbh{F}^{-1}_{(n)}\right).
\end{equation}
\end{widetext}
Carrying out the calculations is simpler than it looks: because of the
structure of eq. (\ref{fullpertexpansion}), each term in the expansion
is proportional to $\Sigma_\bb{p}f_a(N_b,N_p)/N_r^a$, where
$f_a(N_b,N_p)$ is a polynomial in $N_b$ and $N_p$. Terms proportional
to $N_b$ arise from index contractions of type $\mathrm{tr}(\bb{\Psi
  C})=N_b$, which come from symmetrization terms of type
$\Psi_{\{I}\Psi_{J\}}$. Symmetrization factors of type
$\Psi_{\{I}\Psi_J\Psi_{K\}}$ or $\Psi_{\{I}\Psi_{J\}}\Psi_{\{K}\Psi_{L\}}$,
on the other hand, give rise to contractions of type
$\mathrm{tr}(\bb{\Psi
  C})\mathrm{tr}(\bb{\bb{F}\bb{F}^{-1}})=N_bN_p$. Moreover, we know
that, at every order $O(1/N_r^a)$, $f_a(N_b,N_p)$ has to be
proportional to $N_b-N_p$, because it must vanish when $N_b=N_p$. The
reason for this is that if the feature derivative matrix $\bb{d}'_0$
is square and invertible (which it should be in absence of
degeneracies), then eq. (\ref{estimatorcovariance}) reduces to

\begin{equation}
\h{\Sigma}_\bb{p} = (\bb{d}'_0)^{-1}\bb{C}(\bb{d}'^T_0)^{-1}.
\end{equation} 
Every trace of the noise is gone, hence powers of $1/N_r^a$ must not
appear at any order if $N_b=N_p$. Armed with the knowledge of the
above considerations, we can compute the expectation value of
eq. (\ref{fullpertexpansion}) at second, third and fourth order in
$\delta\bbh{\Psi}$, keeping the terms that are at most
$O(1/\nu^2)=O(1/N_r^2)$. When the combinatorial factors that arise
from the expansion of eq. (\ref{fullpertexpansion}) are properly
computed and the expectation values over the inverse Wishart
distribution are taken according to
eqs. (\ref{firstmoment}--\ref{fourthmoment}), the results take the form

\begin{widetext}
\begin{equation}
\label{expansionbreakdown}
\begin{cases}
\begin{displaystyle}
(\delta\bbh{\Psi})^2 \rightarrow \Sigma_\bb{p}\frac{\gamma(N_b-N_p)}{(\gamma-1)(2\gamma+1)} = \Sigma_\bb{p}\left[\frac{N_b-N_p}{N_r}+\frac{(N_b-N_p)(N_b+3)}{N_r^2}\right]
\end{displaystyle} \\ \\

\begin{displaystyle}
(\delta\bbh{\Psi})^3 \rightarrow -4\Sigma_\bb{p}\frac{(N_b-N_p)(1+N_p)}{N_r^2}
\end{displaystyle} \\ \\

\begin{displaystyle}
(\delta\bbh{\Psi})^4 \rightarrow 3\Sigma_\bb{p}\frac{(N_b-N_p)(1+N_p)}{N_r^2}.
\end{displaystyle}

\end{cases}
\end{equation}
\end{widetext}
When the results from eq. (\ref{expansionbreakdown}) are summed, eq. (\ref{quarticdegradation}) immediately follows.

\clearpage
\newpage
\section*{Appendix B: negative effective dimensionality}
\label{appendixB}

The goal of this appendix is to give a justification for why the effective dimensionality $D$ that appears in eq. (\ref{ourscaling}) can be negative in some cases. When we use the same simulation set to estimate $\bb{C},\bbh{\Psi}$, eq. (\ref{estimatorcovariance}) reduces to the inverse Fisher estimator $\h{\Sigma}_\bb{p}=\bbh{F}^{-1}$. At second order in the $\Psi$ fluctuations this becomes

\begin{equation}
\h{\Sigma}_\bb{p} = \bb{F}^{-1} + \bb{F}^{-1}\left(- \delta{\bbh{F}}+ \delta{\bbh{F}}\bb{F}^{-1}\delta{\bbh{F}}\right)\bb{F}^{-1}  
\end{equation}  
If the biased estimator for $\bbh{\Psi}$ is used, we can use eqs. (\ref{firstmoment}--\ref{secondmoment}) at order $O(1/\nu)$ to compute 
\begin{widetext}
\begin{equation}
\label{negativeD}
\langle\h{\Sigma}_\bb{p}\rangle = \Sigma_\bb{p}\left(1-\frac{N_b+1}{N_r}+\frac{1+N_p}{N_r}\right) = \Sigma_\bb{p}\left(1+\frac{N_p-N_b}{N_r}\right).
\end{equation}
\end{widetext}
We immediately see that the coefficient of $1/N_r$ is negative, because $N_b>N_p$. This is the result shown in eq. (\ref{mockscalinguncorrected}). If the bias correction for $\bbh{\Psi}$ is applied, the first order terms $\delta{\bbh{F}}$ average to 0, and we are left with only the last term in the sum eq.~(\ref{negativeD}), which immediately yields eq. (\ref{mockscalingcorrected}).

\label{lastpage}
\end{document}